\documentclass[preprint,aps,floatfix,superscriptaddress,prb]{revtex4}

\usepackage{amsmath, amsthm, amssymb, graphicx}
\usepackage{color}
\usepackage{multirow}
\usepackage{ulem}
\usepackage{float}
\usepackage{bm}
\usepackage{subfiles}
\usepackage{siunitx}
\usepackage{booktabs}
\usepackage{array}

\begin{document}
\setlength{\tabcolsep}{14pt}

\title{Skyrmion relaxation dynamics in the presence of quenched disorder}
\author{Barton L. Brown}
\affiliation{Department of Physics, Virginia Tech, Blacksburg, VA 24061-0435, USA}
\affiliation{Center for Soft Matter and Biological Physics, Virginia Tech, Blacksburg, VA 24061-0435, USA}
\author{Uwe C. T\"{a}uber}
\affiliation{Department of Physics, Virginia Tech, Blacksburg, VA 24061-0435, USA}
\affiliation{Center for Soft Matter and Biological Physics, Virginia Tech, Blacksburg, VA 24061-0435, USA}
\author{Michel Pleimling}
\affiliation{Department of Physics, Virginia Tech, Blacksburg, VA 24061-0435, USA}
\affiliation{Center for Soft Matter and Biological Physics, Virginia Tech, Blacksburg, VA 24061-0435, USA}
\affiliation{Academy of Integrated Science, Virginia Tech, Blacksburg, VA 24061-0563, USA}
\date{\today}

\begin{abstract}
Using Langevin molecular dynamics simulations we study relaxation processes of interacting skyrmion systems with and without quenched disorder. Using the typical diffusion length as the time-dependent length characterizing the relaxation process, we find that clean systems always display dynamical scaling, and this even in cases where the typical length is not a simple power law of time. In the presence of the Magnus force, two different regimes are identified as a function of the noise strength. The Magnus force has also a major impact when attractive pinning sites are present, as this velocity-dependent force helps skyrmions to bend around defects and avoid caging effects. With the exception of the limit of large noise, for which dynamical scaling persists even in the presence of quenched disorder, attractive pinning sites capture a substantial fraction of skyrmions which results in a complex behavior of the two-time auto-correlation function that is not reproduced by a simple aging scaling ansatz.
\end{abstract}

\maketitle

\section{introduction}

Magnetic skyrmions are particle-like spin textures of nanometer size which exist in certain chiral magnets \cite{Nag13,Muh09,Yu10}. In systems without disorder, skyrmions form a triangular lattice due to the long-range repulsive interactions between them, similar to vortices in type-II superconductors. These spin textures have shown great promise for spintronics applications such as racetrack memory devices \cite{Fer13} and logic gates \cite{Zha15} due to their stability, small size, and the low current densities required to move them compared to typical ferromagnetic domain walls \cite{Jon10,Yu12,Iwa13,Lin13}. However, unlike vortices, the non-dissipative Magnus force, which acts perpendicular to the particle velocity, strongly affects the dynamics of moving skyrmions \cite{Nag13,Sch12,Iwa13,Bla94}. In driven systems without disorder the Magnus force causes skyrmions to move at a constant angle with respect to the drive known as the skyrmion Hall angle \cite{Nag13}. In addition, the relaxation dynamics of skyrmion systems without disorder is heavily influenced by the interplay of the Magnus force, the repulsive skyrmion-skyrmion interactions, and the strength of thermal white noise \cite{Bro18}. In general, a strong Magnus force tends to accelerate the relaxation process towards the steady state.

Recently the community started to pay increased attention to the interactions of individual skyrmions with defects \cite{Iwa13,Han16,Liu13,Mul15,Sam13,Kim17,Leg17,Sto17,Sou17,Kos18}. Motivated by the observation that pinning due to material imperfections affects the mobility of skyrmions, different pinning mechanisms were studied theoretically through micro-magnetic simulations of the Landau--Lifshitz--Gilbert equation as well as through Monte Carlo simulations of effective Heisenberg models. Possible defects in magnetic materials that may affect skyrmion motion include locally enhanced magnetic exchange \cite{Liu13}, vacancies \cite{Mul15}, areas with higher magnetic anisotropy \cite{Iwa13,Sam13}, as well as magnetic grains with varying anisotropy \cite{Kim17,Leg17}. Whereas some mechanisms yield attractive pins, others, such as, e.g., regions with enhanced magnetic anisotropy, result in repulsive interactions between defects and skyrmions. Extended one-dimensional defects have been discussed in the context of race tracks \cite{Woo16,Fer13,Kis11,Iwa13,Sam13,Iwa13a,Pur15,Mul17} envisioned as a possible skyrmion-based magnetic storage device.

Collective properties of driven skyrmions have been investigated both experimentally and theoretically. The skyrmion Hall effect has been observed directly \cite{Jia17,Lit17} and described mathematically through a model of driven particle-like skyrmions in a disordered environment \cite{Rei15,Rei16,Dia17} that yields a Hall angle increasing linearly with increasing drive, as observed in experiments. This success of the particle model, in conjunction with the recently observed realignment of skyrmion clusters driven by defect rearrangements \cite{Pol17}, supports the proposed description of skyrmion systems as interacting particle-like objects subject to the Magnus force and to thermal noise \cite{Lin13,Rei16}. Avalanches in driven skyrmions interacting with random pinning have also been investigated \cite{Dia18}. The study of driven interacting skyrmions with pinning has revealed the existence of different dynamic phases and various non-equilibrium phase transitions \cite{Rei18,Bro19}.

Skyrmion systems also share many features with glassy systems. The collective pinning of a skyrmion system due to strong attractive defect sites yields a skyrmion glass that can be investigated in terms of replica field theory \cite{Hos18}.

Recent experimental investigations have revealed that thermal fluctuations have a strong impact on the motion of (non-interacting) skyrmions \cite{Mil18,Zha19}.
It is then natural to ask how the interplay of thermal fluctuations with the Magnus force, the skyrmion-skyrmion interactions, and disorder influences
the collective dynamics of interacting skyrmions. Some partial answers to this question have been provided through the investigation of the steady-state
properties (skyrmion Hall effect) of driven interacting skyrmions. However, as it is the case for many interacting systems, new phenomena are expected
away from stationarity when the system, initially prepared in a non-equilibrium state, relaxes to the steady state. As we discuss in the following,
skyrmion systems show different dynamical regimes, depending on the relative strengths of the Magnus force, the defect pinning strength  and the thermal noise.

In previous work \cite{Bro18}, we showed that the relaxation dynamics of skyrmion systems without disorder can be described through two different aging scaling regimes \cite{Hen10}: one in which the noise dominates the dynamics and the skyrmions behave similarly to non-interacting random walkers; the other in which cooperation between the Magnus force and the repulsive skyrmion-skyrmion interaction leads to the accelerated formation of the triangular lattice and the slow decay of correlations. As was stated previously, the addition of quenched disorder can lead to more complex motion in these systems. In this work, we investigate these phenomena in more detail and study the effects of quenched disorder in the form of randomly placed attractive pinning sites on the relaxation dynamics of many mutually interacting skyrmions. This investigation is carried out through the use of a particle-based model derived from the Landau--Lifshitz--Gilbert equation using Thiele's approach, which is valid in the low-density regime where skyrmions can be treated as point-like particles \cite{Thi73,Nag13,Lin13,Jia17,Pol17}.

The paper is organized in the following way. In Section II we present the model and discuss details of the numerical simulations as well as the measured quantities. Section III compares different types of pinning potentials and shows that the relaxation dynamics of interacting skyrmion systems is essentially unchanged when replacing one type of pins by another.
In Section IV we present our main results. Using the time-dependent diffusion length as our scaling variable, we discuss the extent to which the presence of attractive defects changes the dynamical scaling properties. In the limit of weak thermal noise, skyrmions are easily captured by pinning sites, which yields caging effects and destroys dynamical scaling. We briefly discuss the case of repulsive pinning sites before concluding in Section V.

\section{Simulation details and measured quantities}
In our simulations, we consider $N_{s}=472$ skyrmions interacting on a two-dimensional domain of size $\frac{2}{\sqrt{3}}64\times 64$ with periodic boundary conditions. The number of skyrmions was chosen so that, if the particles were laid out on a grid of unit squares, $\sim$10\% of the domain would be occupied. We model the quenched disorder as $N_{p}=2N_{s}$ randomly placed non-overlapping pinning sites. In the particle-based model presented first in Ref. \cite{Lin13} the Langevin equations of motion are
\begin{equation}
\eta \bm{v}_{i} = \beta\hat{\bm{z}}\times\bm{v}_{i} + \bm{F}^{s}_{i}+\bm{F}^{p}_{i}+\bm{f},
\label{eq_1}
\end{equation}
where $\bm{v}_{i}$ is the velocity of the $i$th skyrmion, $\eta$ is the damping coefficient which aligns the velocity to the net force, and $\beta$ quantifies the strength of the non-dissipative Magnus force which gives the velocity a component perpendicular to the net force. The Magnus force can lead to looping trajectories which allow skyrmions to avoid obstacles and pinning centers in certain cases. Figure 1 qualitatively illustrates the different types of paths skyrmions take during relaxation from a random configuration in the presence of quenched disorder depending on the Magnus force. We impose the constraint $\eta^{2}+\beta^{2}=1$ and consider systems of skyrmions with the Magnus force such that $\beta/\eta=5$ unless specified otherwise \cite{Bro18}. We compare results obtained for this system with those that result from simulating a system of interacting skyrmions without the Magnus force, with $\beta = 0$. The second term on the right-hand side is the repulsive skyrmion-skyrmion interaction which we model as $\bm{F}^{s}_{i}=\sum_{i\neq j}K_{1}(r_{ij})\hat{\bm{r}}_{ij}$ where $K_{1}(x)$ is the modified Bessel function of the second kind, which falls off exponentially with large $x$, and $\hat{\bm{r}}_{ij}=\frac{\bm{r}_{i}-\bm{r}_{j}}{|\bm{r}_{i}-\bm{r}_{j}|}$ is a unit vector where $\bm{r}_{i}$ is the location of the $i$th skyrmion and $r_{ij}=|\bm{r}_{i}-\bm{r}_{j}|$. The third term is the force on skyrmion $i$ due to pinning sites which we assume to be simple harmonic traps for the main part of this work given by $\bm{F}^{p}_{i}=-\sum_{j}^{N_{p}}(F_{0}^{p}\,|\bm{r}_i-\bm{r}_j^p|/r_{p})\hat{\bm{r}}_{ij}$ where $F_{0}^{p}$ is the maximum force the pinning site can exert on a skyrmion, $r_{p}$ is the pinning radius, $\bm{r}_j^p$ is the position of the $j$th pin, and $\hat{\bm{r}}_{ij}=\frac{\bm{r}_{i}-\bm{r}^p_{j}}{|\bm{r}_{i}-\bm{r}^p_{j}|}$ is a unit vector. 
This force only acts on a skyrmion as long as its distance to the pinning center is smaller than the pinning
radius, i.e., provided $\left| \bm{r}_i - \bm{r}^p_j \right| \le r_p$.
In this work, we use the pinning strength $F^{p}_{0}=0.1$ and radius $r_{p}=0.1$ unless otherwise specified. We also briefly discuss the case of purely repulsive defects with $F^{p}_0=-0.1$. In Section III, we compare the fraction of pinned particles and the two-time density auto-correlation function measured in systems with simple harmonic pins to systems with smooth hyperbolic-tangent type pins and find that the results agree almost exactly under the appropriate choice of pinning radius. The last term in Eq. (\ref{eq_1}) represents thermal white noise with strength characterized by $\sigma$ obeying the following relations: $\langle f_{\mu}(t)\rangle=0$ and $\langle f_{\mu}(t)f_{\nu}(t')\rangle=\sigma\delta_{\mu\nu}\delta(t-t')$. We numerically solve the equations of motion using a standard fourth-order Runge--Kutta method with a time step of $\Delta t=0.1$.

\begin{figure} 
  \begin{minipage}[b]{0.33\linewidth}
    \centering
    \includegraphics[width=1\linewidth]{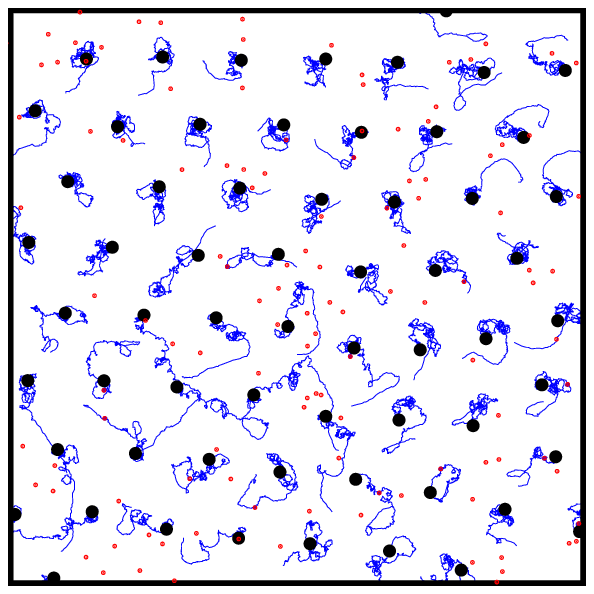}
    \vspace{0ex}
  \end{minipage}
  \begin{minipage}[b]{0.33\linewidth}
    \centering
    \includegraphics[width=1\linewidth]{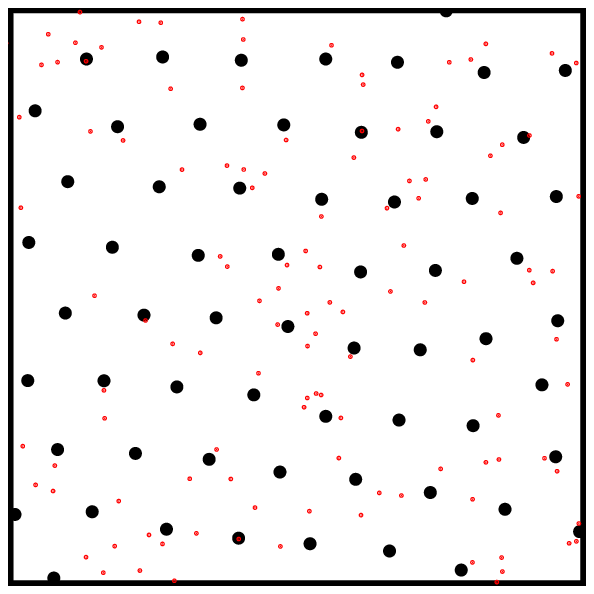}
    \vspace{0ex}
  \end{minipage}
  \begin{minipage}[b]{0.33\linewidth}
    \centering
    \includegraphics[width=1\linewidth]{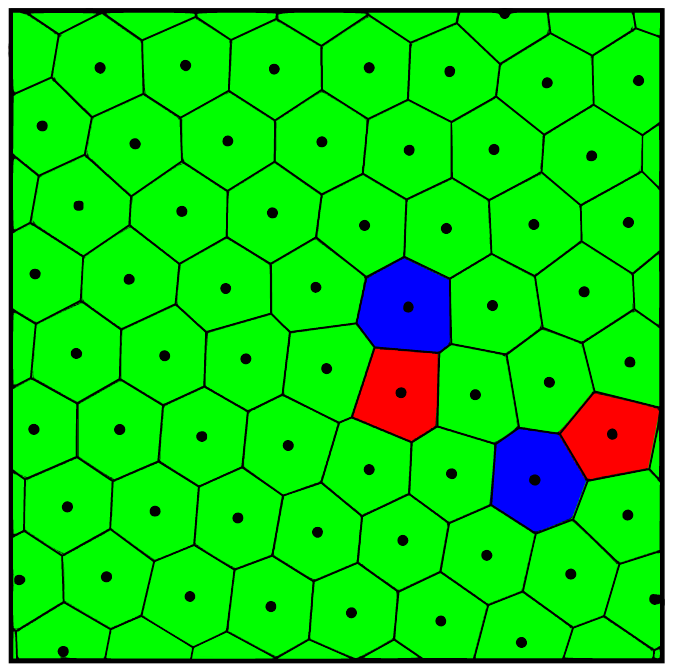}
    \vspace{0ex}
  \end{minipage}
  \begin{minipage}[b]{0.33\linewidth}
    \centering
    \includegraphics[width=1\linewidth]{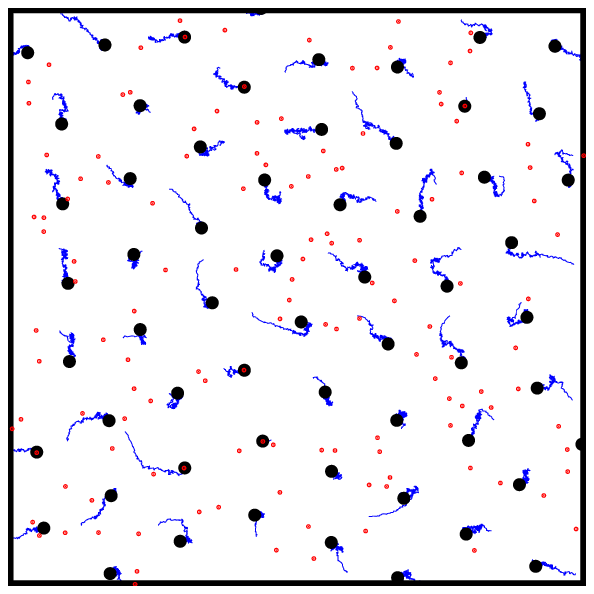}
    \vspace{0ex}
  \end{minipage}
  \begin{minipage}[b]{0.33\linewidth}
    \centering
    \includegraphics[width=1\linewidth]{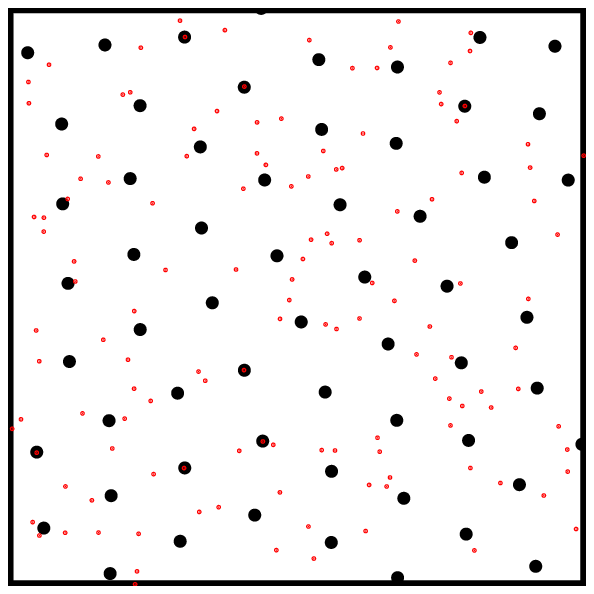}
    \vspace{0ex}
  \end{minipage}
  \begin{minipage}[b]{0.33\linewidth}
    \centering
    \includegraphics[width=1\linewidth]{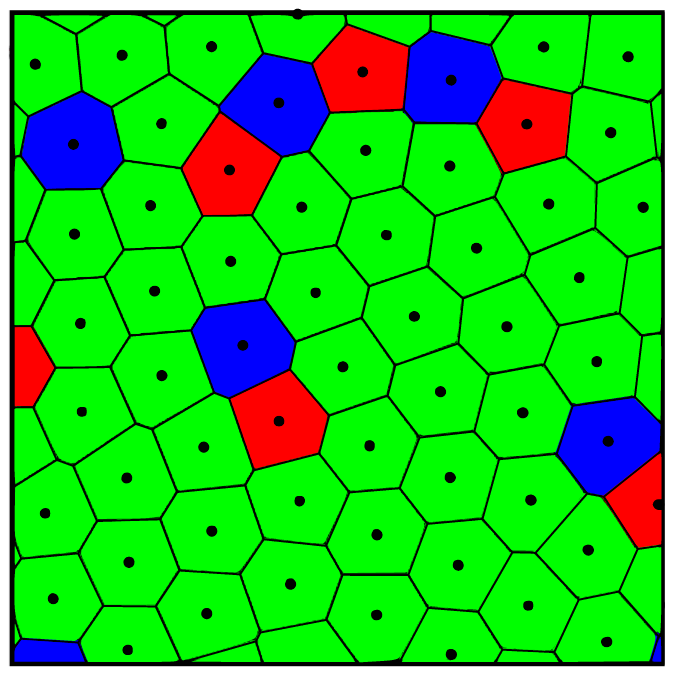}
    \vspace{0ex}
  \end{minipage}
  \label{figure1}
  \caption{Selected regions from two systems: with the Magnus force (top) and without (bottom). The skyrmions (black circles) start from a disordered state and relax in the presence of the attractive pinning centers (red circles). (Left) The paths of the particles (blue lines) from $t=10$ to $t=10,000$ are shown. In the case of the Magnus force, the trajectories bend and loop, avoiding pinning centers. (Middle) The location of each skyrmion and pinning site at $t=10,000$. (Right) Voronoi tessellations of the selected regions at $t=10,000$. Regions in green have six edges, whereas regions in blue and red have seven and five edges, respectively. The skyrmion lattice has fewer defects in the presence of the Magnus force, which is due to the fact
that the Magnus force accelerates the relaxation of the system. In both simulations, the noise strength is set to $\sigma=0.1$ which accounts for the fluctuations in the paths.}
\end{figure}

Initially, $N_{s}$ skyrmions are randomly placed in the domain. In the absence of pinning sites and thermal noise, the skyrmions relax into a triangular lattice configuration. However, in the presence of quenched disorder skyrmions can become trapped leading to distortions in the lattice and caging effects which can further impede the motion. In the case of strong thermal noise, the skyrmions are able to escape from pinning sites and essentially undergo a random walk. The system therefore does not reach the triangular lattice configuration but a disordered steady state. During the relaxation process we measure the fraction of pinned particles as a function of time, the average mean-square displacement, and the two-time density auto-correlation function, in simulations with and without the Magnus force.

Having placed the $N_{s}$ skyrmions randomly on the domain, the system begins to relax starting at time $t=0$ towards the steady state which depends on the strength of the pinning sites, the noise, and the Magnus force. We measure the fraction of pinned particles $f_{p}$ as
\begin{equation}
f_{p}=\Bigg\langle \frac{1}{N_{s}}\sum_{i=1}^{N_{s}}\sum_{j=1}^{N_{p}} \Theta ( r_{p}-|r_i-r_j^p| ) \Bigg\rangle,
\end{equation}
where $\Theta$ denotes the Heaviside step function and the brackets $\langle \ldots \rangle$ indicate an average over different realizations of the noise,
different
pinning site configurations, and different initial skyrmion positions. During relaxation, $f_{p}$ grows monotonically before saturating. In the limit of strong thermal fluctuations, the saturation value is very close to zero and displays large fluctuations as skyrmions can easily escape pinning centers. 

In order to quantitatively investigate the relaxation process and aging phenomena we look to the two-time density auto-correlation function which has proven useful in understanding the dynamics in this system in the absence of pinning sites \cite{Bro18}. As the skyrmions relax from their random initial positions due to the repulsive interactions and attractive pinning interactions, the correlation function decays gradually from unity. We follow Ref. \cite{Ple11} and measure the correlation function for a specific waiting time $s$ by designating a circular region of radius $r_{a}$ centered at the location of each skyrmion at time $t=s$. 
We find that our results do not depend on the choice of $r_{a}$ as long as it is 
much smaller than the long range of the repulsive skyrmion-skyrmion interactions and not
much larger than the pinning radius $r_p$.
The force between skyrmion $i$ and $j$ becomes sufficiently close to zero to be negligible at $r_{ij}\approx10$. In this work we chose $r_{a}=0.1$. This value is larger than that in Ref. \cite{Bro18} as we want to have $r_a \approx r_p$. Then, for $t\ge s$ we define the occupation number $n_{i}(t)$ which is equal to $1$ if the $i_{th}$ skyrmion is within the designated circular region at time $t$ and $0$ otherwise. The two-time density auto-correlation function is then given by
\begin{equation}
C(t,s) = \Bigg\langle \frac{1}{N_{s}}\sum_{i=1}^{N_{s}}n_{i}(t)n_{i}(s)\Bigg\rangle.
\end{equation}
We measure $C(t,s)$ for several different values of the waiting time $s$ during the relaxation process. 

In many ordering systems the correlation function exhibits a simple aging scaling behavior \cite{Hen10} 
\begin{equation}
C(t,s)=\left(L(s)\right)^{-b}f_{C}\left( L(t)/L(s) \right),
\label{ag_eq}
\end{equation}
where $b$ is called the aging exponent. $L(t)$ is the value at time $t$ of a characteristic time-dependent length so that the scaling function $f_{C}$ only depends on the ratio $L(t)/L(s)$ of this typical length measured at two different times. In cases where this length grows algebraically with time (which is for example the case for systems undergoing domain coarsening) the correlation takes on the simpler form 
\begin{equation}
C(t,s)=s^{-\tilde{b}} \tilde{f}_C(t/s)
\label{ag_eq2}
\end{equation}
in the aging scaling regime \cite{Hen10}. In our previous analysis \cite{Bro18}, done without having at our disposal a good quantity for calculating the typical time-dependent length, we observed two regimes of interest when assuming the simpler scaling form (\ref{ag_eq2}). In the case of strong thermal fluctuations, the Magnus force serves to enhance the effects of the thermal noise and the aging exponent $\tilde{b}$ is positive. In contrast, when the noise is weak the Magnus force accelerates the formation of the triangular lattice and the exponent is negative. We revisit this analysis in the following where we use as time-dependent length the square root of the average mean-square displacement, $L(t) = \sqrt{r^2(t)}$, with 
\begin{equation}
r^{2}(t)=\Big\langle \frac{1}{N_{s}}\sum_{i}^{N_{s}}\Big|\bm{R}_{i}(t)-\bm{R}_{i}(0)\Big|\Big\rangle,
\label{eq_r2}
\end{equation}
with the absolute coordinates $\bm{R}_{i}$ in which the periodic boundary conditions are not applied. Similar to $f_{p}$, the typical diffusion length $L(t)$ saturates after a period of monotonic increase in systems with strong caging effects. 

\section{Pinning potential comparison}

Before discussing the relaxation processes of interacting skyrmions in the presence of attractive pinning sites, it is important to understand whether different choices of pining potentials affect the dynamic properties of our system.
We consider in this section two types of radially symmetric pinning potentials: one in the form of a hyperbolic tangent which has a smooth transition at the edge of the pinning region given by 
\begin{equation}
U_{t}(r) = -\frac{1}{5}F^{p}_{0}r^t_p\Bigg[1-\textrm{tanh}\Bigg(5\frac{r-r^t_p}{r^t_p}\Bigg)\Bigg]~, \end{equation}
and a simple harmonic potential, with 
\begin{equation}
U_h(r) = (F^{p}_{0}/2r^h_{p})\left( r^{2} - (r^h_{p})^{2}\right)~. 
\end{equation}
Here $r = \left| \bm{r} - \bm{r}^p \right|$ is the distance between skyrmion position $\bm{r}$ and pinning site $\bm{r}^p$, and the potentials are zero
for distances larger than the pinning radius $r_p^t$ or $r_p^h$.
The attractive pinning force is then calculated in the standard manner, $\bm{F}^{p}=-\frac{\partial U(r)}{\partial r}\hat{\bm{r}}$.

\begin{figure}
    \centering
    \includegraphics[width = 0.7\linewidth]{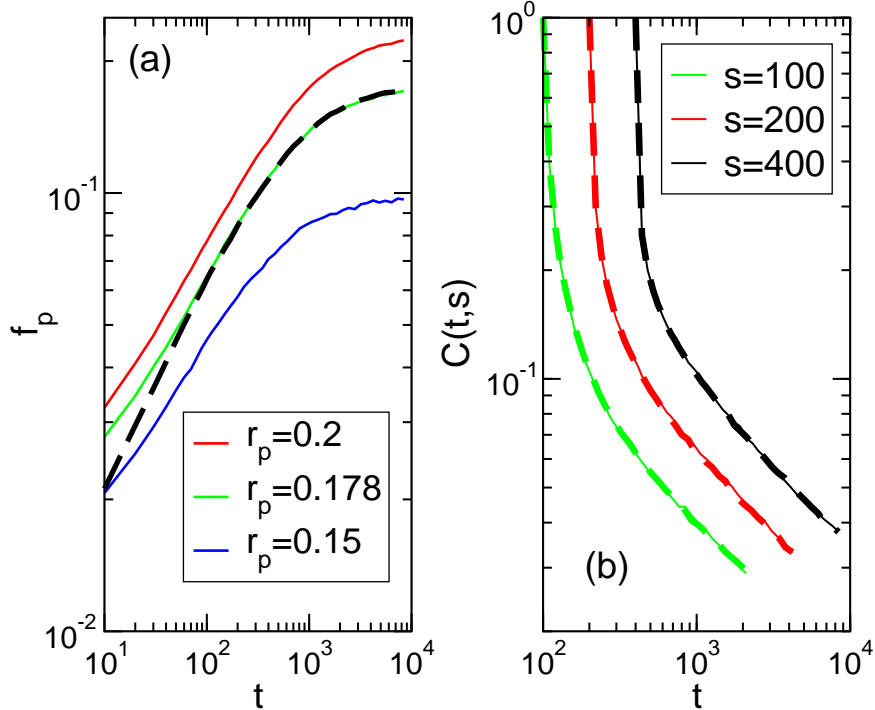}
    \caption{(a) The fraction of pinned particles as a function of time for the harmonic type pinning potential with radius $r^h_{p}=0.1$ (black dashed line) and the hyperbolic-tangent type potential for various values of $r^t_{p}$ (solid lines). The data for each curve was generated by averaging over 200 realizations with the appropriate pin type and size. The pinning strength $F^{p}_{0}=0.1$ was used for both pin types. The radius is varied from $r^t_p=0.15$ to $0.2$ for the hyperbolic tangent pins and we find that a value of $r^t_p=0.178$ reproduces the pinned fraction of the harmonic type pins very well. (b) The two-time density auto-correlation as a function of time measured in simulations using the hyperbolic-tangent type pins with $r^t_p=0.178$ (solid lines) and the harmonic type pins with $r^h_p=0.1$ (dashed lines) for several waiting times $s$. In all cases, the noise strength $\sigma=0.1$ was used.}
    \label{fig2}
\end{figure}

In order to understand how the pin potential can change the relaxation dynamics, we measure the fraction of pinned particles and the two-time density auto-correlation function for simulations in which pinning sites of either potential $U_{t}$ or $U_{h}$ are used. In Fig. \ref{fig2}a we compare for $f_{p}$ results obtained in simulations using the harmonic type pins with $r^h_{p}=0.1$ to those obtained in simulations using hyperbolic-tangent type pins with several different values of $r^t_{p}$. We find that for an appropriate choice of $r^t_{p}\approx 0.178$, the fraction of pinned particles does not depend on the functional form of the pinning potential after a transient time. This suggests that the relaxation dynamics are not inherently dependent on the shape of the radially symmetric well. In Fig. \ref{fig2}b, we show the two-time density correlation function for different values of the waiting time $s$ using the harmonic type pins with $r^h_{p}=0.1$ (dashed lines) and the hyperbolic-tangent type pins with $r^t_p = 0.178$ (solid lines). The results are in excellent agreement with each other, implying that the relaxation dynamics in our system do not strongly depend on the shape of the radially symmetric wells. 
In the remaining sections, we focus on the case of harmonic pinning sites. As no confusion is possible we drop in the following the index $h$ and use $r_p$ for the pinning radius.

\section{Results}

In the absence of pinning sites the Magnus force has the effect of accelerating the relaxation process \cite{Bro18}. However, the details of this accelerated relaxation depend on the interplay between the thermal noise and the Magnus force. In the thermal-noise dominated regime the Magnus force further enhances the disordering effect of the noise. On the other hand, in the Magnus-force dominated regime one observes a complicity between the repulsive skyrmion-skyrmion interactions and the Magnus force that results in an accelerated ordering in the form of an (approximate) triangular lattice. 

As already mentioned, the presence of a time-dependent typical length allows a more appropriate scaling analysis where this length is used in the argument of the scaling functions. Using the square root of the average mean-square displacement (\ref{eq_r2}) as our characteristic length $L(t)$, we first revisit the scaling behavior of clean systems before extending this analysis to interacting skyrmions in the presence of quenched disorder in the form of attractive pinning sites.

\begin{figure}
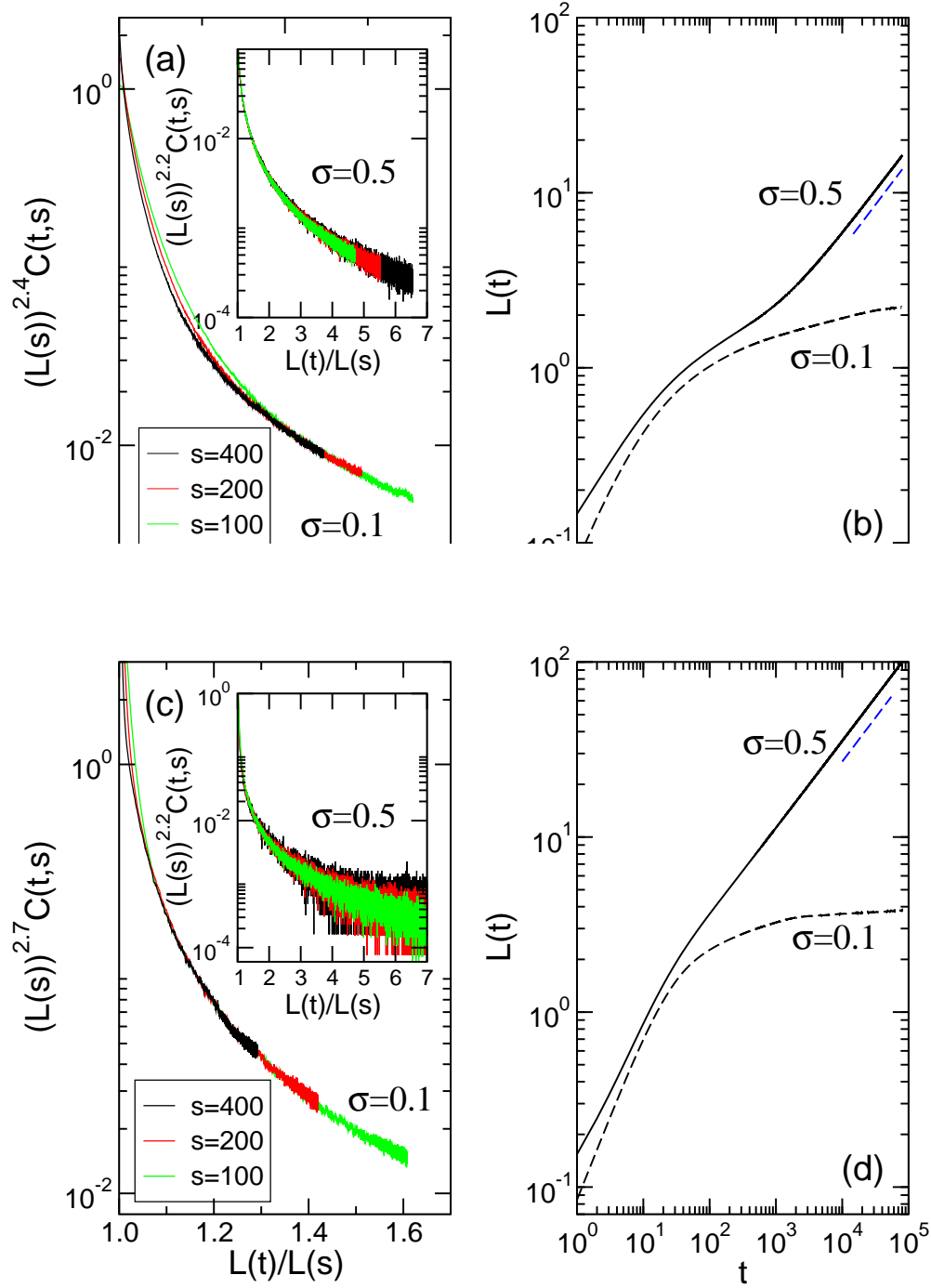

    \centering
    \includegraphics[width=0.8\linewidth]{figure3a.eps}\\
    \includegraphics[width=0.8\linewidth]{figure3b.eps}
    \caption{Scaled two-time density auto-correlation functions (\ref{ag_eq}) and the time-dependent typical diffusion lengths $L(t)$ in the absence of pinning sites: (a,b) without the Magnus force, (c,d) with the Magnus force. For each case results for two different noise strengths are shown. The dashed blue lines in (b,d) indicate a power law with an exponent 0.5. This data result from averages over 700 independent runs.}
    \label{fig3}
\end{figure}

The results of our analysis in the absence of pins are summarized in Fig. \ref{fig3} and in Table \ref{Tab:scaling_no_pins}. In panels Fig. \ref{fig3}a and \ref{fig3}c we show the scaling behavior of the two-time density autocorrelation function without and with the Magnus force, each for two different values of the noise strength: $\sigma = 0.1$ in the main panels and $\sigma = 0.5$ in the insets. As scaling ansatz we choose (\ref{ag_eq}) with the typical diffusion length, displayed in Fig. \ref{fig3}b and \ref{fig3}d for the same cases, as our time-dependent length $L(t)$. For $\sigma = 0.5$ we observe both in the presence and absence of the Magnus force an excellent aging scaling behavior with an exponent $b=2.2(1)$, slightly larger than the value 2 for independent random walkers, the expected behavior for very large noise strengths $\sigma$. Inspection of $L(t)$ reveals that for $\sigma = 0.5$ this length, after an initial transient regime, changes algebraically with time, with the exponent 0.5 of normal diffusion (indicated by the dashed blue lines). In the presence of the Magnus force particles enter this normal diffusion regime earlier than in its absence, which results in a much larger value of the typical diffusion length at the end of our runs. This is of course a manifestation of the fact that in the noise-dominated regime the Magnus force enhances the effects of the noise. Aging scaling is also observed in both cases for weak noise strengths, as shown in Fig. \ref{fig3} for $\sigma = 0.1$. We note that for this value of $\sigma$ the typical length does not vary algebraically with time, but instead approaches a plateau. In agreement with the observation that the Magnus force helps the system for low noise values to relax quickly to the triangular lattice, the typical diffusion length in the presence of the Magnus force increases more strongly than in the absence of this force, and quickly approaches a limiting value. This acceleration of the relaxation towards the hexagonal lattice becomes also apparent as an increase of the value of the exponent $b$, changing from 2.2 to 2.7, see Table \ref{Tab:scaling_no_pins}, when $\sigma$ changes from 0.2 to 0.1. This change reveals the dynamical crossover between the noise-dominated regime for large $\sigma$ and the Magnus-force dominated regime for small $\sigma$ \cite{Bro18}.

\begin{table}                                            
\center                          
  \begin{tabular}{cc||cc}
    \toprule[1.5pt]                                          
    \multicolumn{2}{c||}{without Magnus force} & \multicolumn{2}{c}{with Magnus force}\\
    \hline                                     
    $\sigma$ & $b$  & $\sigma$ & $b$\\
    \hline                                                  
   $\infty$ & 2.0   & $\infty$ & 2.0\\
    0.5 & 2.2(1)          & 0.5 & 2.2(1)\\
    0.2 & 2.3(1)         & 0.2 & 2.2(1)\\
    0.1 & 2.4(1)        & 0.1 & 2.7(1)\\           
    0.05 & 2.4(1)        & 0.05 & 2.85(10)\\           
    0.0 & 2.4(1)        & 0.0 & 2.9(1)\\
    \bottomrule[1.5pt]
  \end{tabular}
  \caption{Values of the aging scaling exponent without pinning sites as function of noise strength $\sigma$.}                                        
  \label{Tab:scaling_no_pins}
  \end{table}

\begin{figure}
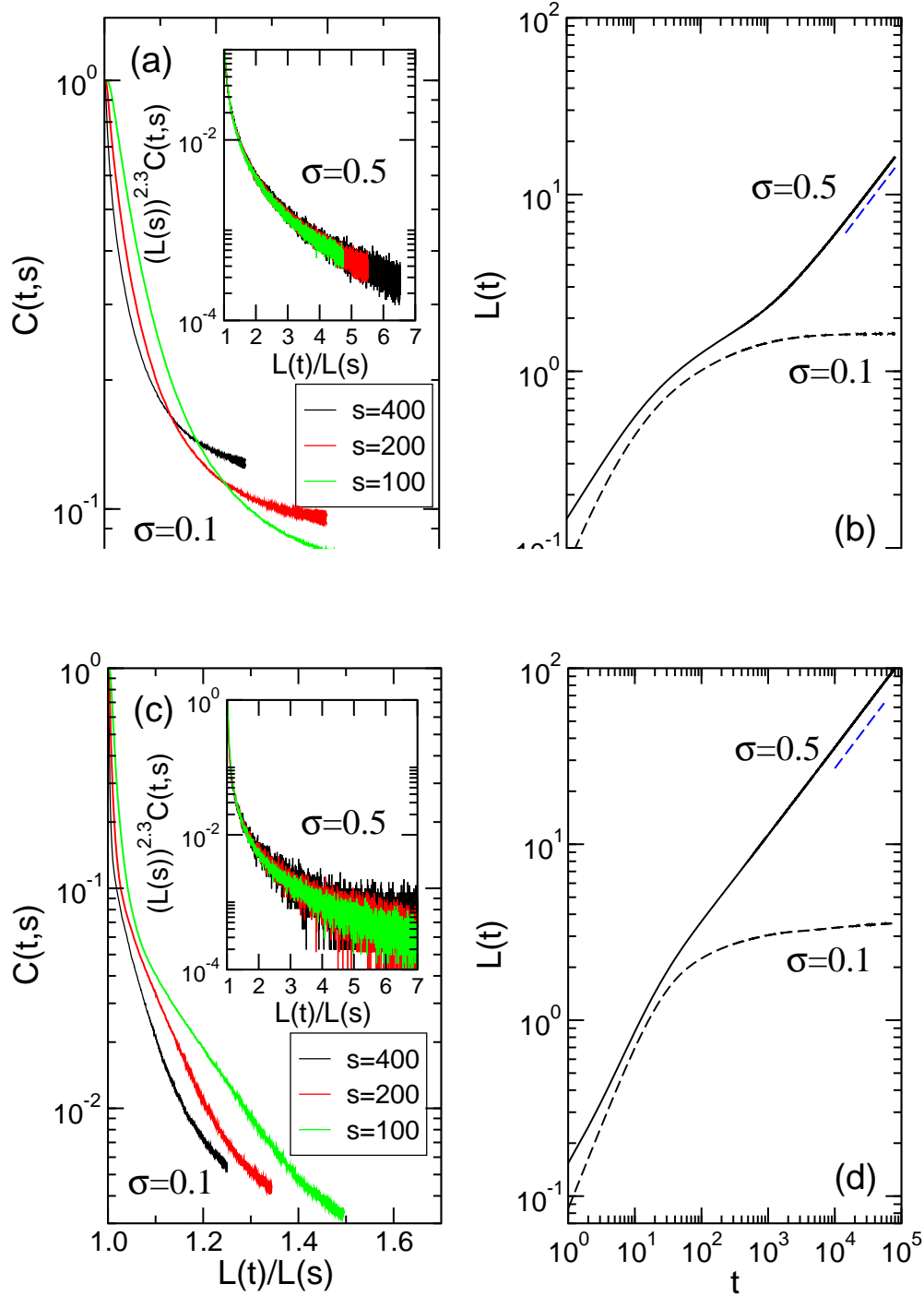

    \centering
    \includegraphics[width=0.8\linewidth]{figure4a.eps}\\
    \includegraphics[width=0.8\linewidth]{figure4b.eps}
    \caption{Scaled two-time density auto-correlation functions (\ref{ag_eq}) and the time-dependent typical diffusion lengths $L(t)$ in the presence of attractive pinning sites: (a,b) without the Magnus force, (c,d) with the Magnus force. For each case results for two different noise strengths are shown. The dashed blue lines in (b,d) indicate a power law with an exponent 0.5. This data was collected from simulations with harmonic traps with $r_{p}=0.1$ and averaged over 700 different realizations of the noise,
initial skyrmion positions, and pinning site configurations.}
    \label{fig4}
\end{figure}

Adding attractive pinning sites does not affect the behavior for large noise strengths, as illustrated in Fig. \ref{fig4} and Table \ref{Tab:scaling_pins}. This changes dramatically for small values of $\sigma$ for which there is an increased probability for a sizable fraction of skyrmions to get stuck at pinning sites. Consequently, the ordering process is impeded, which results in the absence of dynamical scaling, see the data in Fig. \ref{fig4} for $\sigma = 0.1$. In the absence of the Magnus force, aging scaling is not observed for much larger values of $\sigma$ than in the presence of this force. 

\begin{figure}
    \centering
    \includegraphics[width=0.7\linewidth]{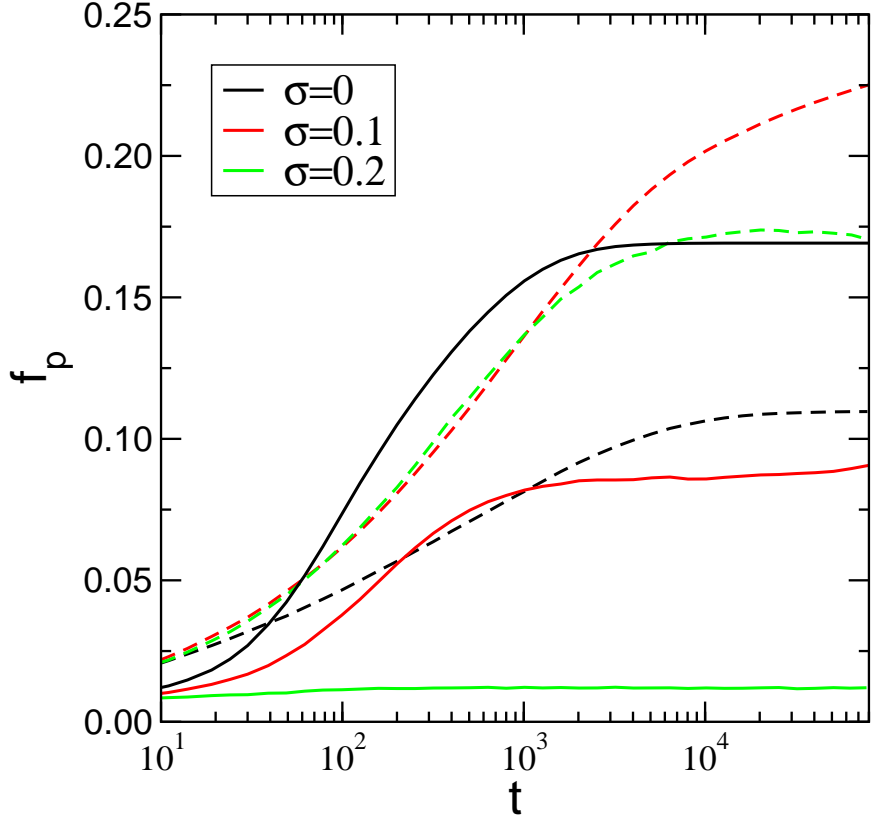}
    \caption{The fraction of pinned skyrmions as a function of time with (solid lines) and without (dashed lines) the Magnus force for different noise strengths. Only attractive pins are present in the system. In the absence of noise, the Magnus force allows skyrmions to find pinning sites more quickly, which increases $f_{p}$ compared to systems without the Magnus force. As the noise strength increases in the presence of the Magnus force, the fraction of pinned skyrmions decreases monotonically. In contrast, without the Magnus force $f_p$ displays a non-monotonic behavior upon enhancing the noise strength: increasing first, as thermal fluctuations allow particles to overcome strong caging effects, thus increasing their mobility, followed by a decrease once the noise strength allows the skyrmions to escape from pinning centers. This data was averaged over 1000 different realizations of the noise, initial skyrmion positions, and pinning site configurations.}
    \label{fig5}
\end{figure}

A more refined view of the interplay of the Magnus force, the noise, and the attractive pins is provided in Fig \ref{fig5}, where we show the time dependence of the fraction of skyrmions that are captured at pinning sites. In systems experiencing the Magnus force (full lines in the figure), this fraction increases monotonically with decreasing noise levels. Skyrmions that are less jolted through thermal noise have a higher probability to be captured by pins. Comparing for a fixed noise strength $\sigma > 0$ the data with and without the Magnus force, one notes that without the Magnus force (dashed lines) a substantially larger fraction of skyrmions are captured. Indeed, the Magnus force provides an additional mobility to the skyrmions that in conjunction with the noise enhances the probability of a skyrmion to escape from a pin. The case with no noise and no Magnus force (dashed black line) is a special one: already a rather small fraction of pinned skyrmions yields (almost) blocked configurations where due to caging effects large numbers of skyrmions have very restricted mobility. This behavior is reminiscent of the freezing dynamics encountered in glass forming systems.
\begin{table}                                                  
\center                          
  \begin{tabular}{cc||cc}
    \toprule[1.5pt]                                          
    \multicolumn{2}{c||}{without Magnus force} & \multicolumn{2}{c}{with Magnus force}\\
    \hline                                     
    $\sigma$ & $b$  & $\sigma$ & $b$\\
    \hline                                                  
   $\infty$ & 2.0   & $\infty$ & 2.0\\
    0.5 & 2.3(1)    & 0.5 & 2.3(1)\\
    0.2 & no scaling   & 0.2 & 2.3(1) \\
    0.1 & no scaling    & 0.1 & no scaling\\           
    0.05 & no scaling    & 0.05 & no scaling\\           
    0.0 & no scaling    & 0.0 & no scaling\\
    \bottomrule[1.5pt]
  \end{tabular}
  \caption{Values of the aging scaling exponent with attractive pinning sites as a function of noise strength $\sigma$.}                                        
  \label{Tab:scaling_pins}
  \end{table}

We also briefly studied the case of purely repulsive defects (for which we change the sign of the pre-factor of the pinning force but leave all other parameters unchanged) for the case with the Magnus force and noise strengths $\sigma = 0$ and $\sigma = 0.2$. Comparing the results obtained for these two cases with the clean system, we note that neither the typical diffusion length nor the auto-correlation function are much affected by the repulsive defects at the studied density, yielding the same exponents for these two quantities. The only notable effect is an increase of the noisiness of the auto-correlation data. Clearly the interacting skyrmions are not much perturbed in their relaxation process when the combined area of all the defects is only $\sim 0.6 \%$ of the total domain. We do expect stronger effects for increasing defect densities, as at some point the skyrmions will no longer be able to settle into a more or less unperturbed triangular lattice.

\section{Conclusion}

Our study of the effects of quenched disorder on systems of interacting skyrmions has yielded new insights into the non-equilibrium relaxation process starting from randomized initial configurations, and the role played by the Magnus force in situations where the effective particle picture is valid. Using the typical diffusion length as the characteristic time-dependent length for the relaxation processes, we have been able to describe in a consistent manner distinct dynamic regimes, and this independently of whether the length varies algebraically with time or not.

Our new analysis reveals for clean systems the existence of aging scaling of the two-time density auto-correlation function for a variety of situations, in the presence and absence of the Magnus force as well as in the presence and absence of noise. Without the Magnus force the value of the aging exponent is basically independent of the noise strength $\sigma$, as shown in Table \ref{Tab:scaling_no_pins}. This is different in the presence of the Magnus force where a rapid change in the value of the exponent indicates a transition from a noise-dominated regime at large values of $\sigma$ to a Magnus-force dominated regime at small noise strengths. This transition has already been observed in our previous work \cite{Bro18}, but our current analysis is in fact considerably more consistent as we use the scaling form (\ref{ag_eq}) that relies on the time-dependent typical length and does therefore not a-priori assume a power-law growth (which we do not actually observe for smaller values of $\sigma$) of the typical length.

Quenched disorder in the form of randomly placed attractive pins has a major effect on the relaxation processes as it limits mobility of the skyrmions and for low noise levels introduces strong caging effects. As a result, dynamical scaling is not encountered anymore in our system, with the exception of very high noise strengths for which the system does not proceed to order into a (distorted) triangular lattice. Interestingly, the fraction of pinned skyrmions shown in Fig. \ref{fig5} reveals that the presence of the Magnus force has a major impact in the low-noise regime. Indeed, by adding a velocity component in the direction perpendicular to the direction of propagation, the Magnus force allows skyrmions to bend around obstacles and pinning centers rendering the dynamics far more robust overall against pinning effects. As a result there are for $\sigma > 0$ far fewer skyrmions pinned in the presence of the Magnus force than in its absence. In the noiseless case ($\sigma=0$), the systems quickly become stuck in a blocked configuration if the Magnus force is not present. A non-vanishing Magnus force allows the skyrmions to overcome strong caging effects and to avoid blocked states, and this even in the absence of any thermal noise.

If instead of attractive pins we implement repulsive defects, our simulations yield results that are very similar to those obtained in the clean system without any defects. Indeed, when the density of defects is low (which is the case in our simulations), the skyrmions are able to avoid the space occupied by the defects and still order in a near-perfect triangular lattice. We expect that more complex features emerge for much larger excluded areas (due to a much larger number of repulsive defects) and plan to study this in more detail in the future.

\begin{acknowledgments}
This research was supported by the US Department of Energy, Office of Basic Energy Sciences, Division of Materials Sciences and Engineering under Grant No. DE-SC0002308.
\end{acknowledgments}

\end{document}